# Characterization of transiting exoplanets by way of differential photometry

**Michael Cowley[1,2] and Stephen Hughes[2]**

[1] Department of Physics and Astronomy, Faculty of Science, Macquarie University, Sydney, NSW 2109, Australia
[2] School of Chemistry, Physics and Mechanical Engineering, Science and Engineering Faculty, Queensland University of Technology, Gardens Point Campus, Brisbane, Queensland 4001, Australia

E-mail: michael.cowley@mq.edu.au

**Abstract**
This paper describes a simple activity for plotting and characterising the light curve from an exoplanet transit event by way of differential photometry analysis. Using free digital imaging software, participants analyse a series of telescope images with the goal of calculating various exoplanet parameters, including its size, orbital radius and habitability. The activity has been designed for a high school or undergraduate university level and introduces fundamental concepts in astrophysics and an understanding of the basis for exoplanetary science, the transit method and digital photometry.

**Introduction**
Since the discovery of the first planet outside of our Solar System in 1992 [1], the study and characterisation of exoplanets has become one of the most dynamic fields of research in astrophysics. Knowledge in the area has grown exponentially and has contributed to an improved understanding of planetary formation and evolution. Most exoplanet discoveries have resulted from observing the stars, as direct observation is hindered by the small amount of light an exoplanet reflects. One such indirect technique, known as the transit method, has successfully yielded more than one hundred discoveries in the past decade. The transit method involves detecting the slight dip in brightness of a host star when an exoplanet passes in front of it, as viewed by an observer (see Fig.1).

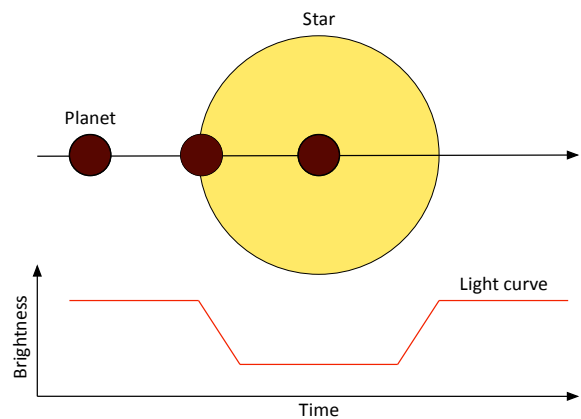

**Figure 1.** Image showing transit of exoplanet in front of a star from left to right, and the resulting dip in the light curve's brightness

In this paper, we outline a simple activity that allows students to analyse a series of telescope images, plot their own transit light curve and calculate a number of exoplanetary system characteristics, including its size, orbit distance and its likelihood for habitability. This is achieved by way of differential photometry, which is an image processing technique used to compare the relative change in brightness between a target and reference stars in an image. The exercise requires access to a computer with spreadsheeting software, internet access and the ability to download and install free digital imaging software. It can easily be completed within 3 hours in a high school or undergraduate university classroom setting by the students themselves.

**Practical Activity**
Thanks to a number of observatories sharing data publicly, time-series images of numerous host stars are readily available to download from a number of sources. The activity described in this paper involves detecting, analysing and chracterising the exoplanet, WASP-2b, which was originally discovered by the SuperWASP project in 2006 by way of the transit method [2]. The data, comprised of 10 time-sequenced images of WASP-2, was sourced from the Las Cumbres Observatory Global Telescope Network [3].

**Image Analysis**
To analyse the time-sequenced images, we used the freely available AstroImageJ by the University of Louisville (see Fig.2). AstroImageJ is a modified version of the image processing software, ImageJ [4]. AstroImageJ comes with a detailed manual, but we suggest educators provide students with a basic overview for the activity (www.astro.louisville.edu/software/astroimagej/).

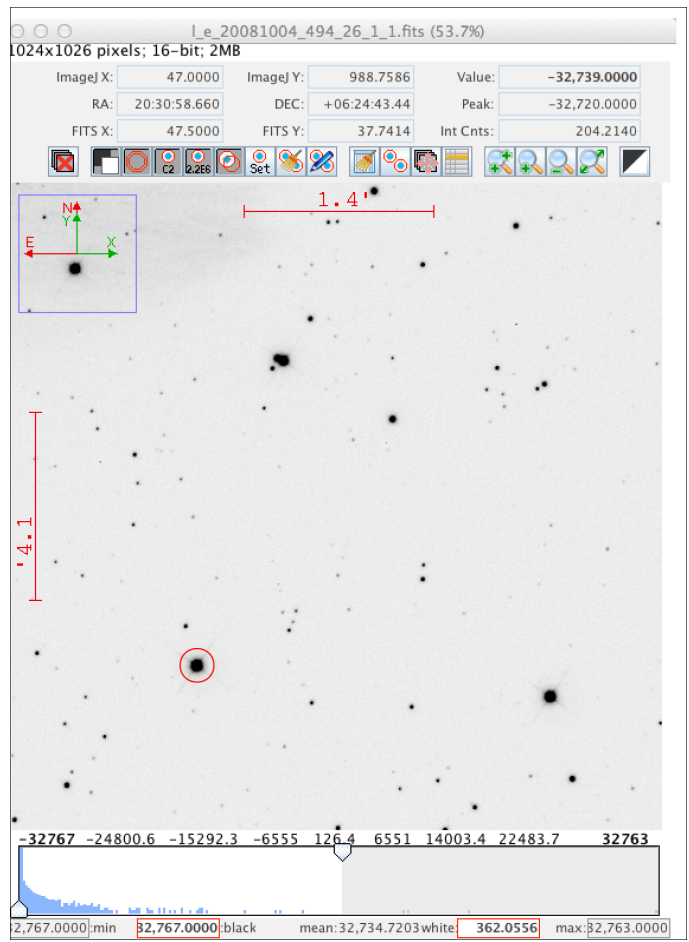

**Figure 2.** Sample frame from WASP-2b (circled) image in AstroImageJ

As mentioned above, this activity involves measuring the brightness of stars to look for any changes that may be consistent with a transit event. Unfortunately, the brightness of each star will appear to fluctuate from frame to frame. When images are photographed with a long exposure, fluctuations are averaged and the final image will result in fuzzy stars. Known as astronomical seeing, this phenomenon only impacts ground-based observations and is a result of the turbulent mixing in Earth's atmosphere. When measuring the total brightness of individual stars, two different areas of light are considered – light from the star (smeared due to astronomical seeing) and background light (such as moonlight). The aperture photometry tool in AstroImageJ allows for each of these areas to be assigned as a circular annulus, as shown in Fig.3. This tool adds all light that has been smeared by the seeing quality, and subtracts all light from the background.

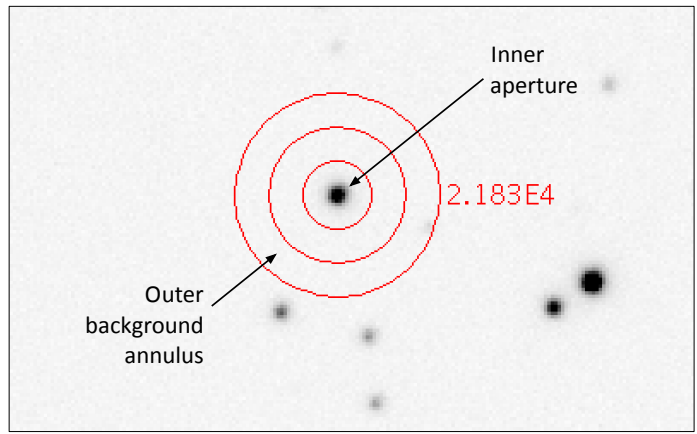

**Figure 3.** The inner source aperture and the outer background annulus in AstroImageJ

For measurements on the WASP-2b images, it is recommended that the radius of the inner aperture be set to 18, the inner background radius to 40, and the outer background radius to 60. Measurements should be conducted on the host star and a number of reference stars to correct for variations induced by factors other than an exoplanet transit, such as eclipsing binaries and variable stars. AstroImageJ supports multi-aperture photometry, which allows for automatic photometric measurements on multiple objects within an image over a sequence. However, as there are only 10 images in this sample, it is recommended each image be inspected individually. Upon selecting a star, results will be displayed in a table of measurements, where the brightness is listed as "Source-Sky". The time and brightness of the target and at least one reference star should be recorded for each image. Plotting the measurements of the host divided by the reference star over time should result in a transit light curve, as shown in Fig.4.

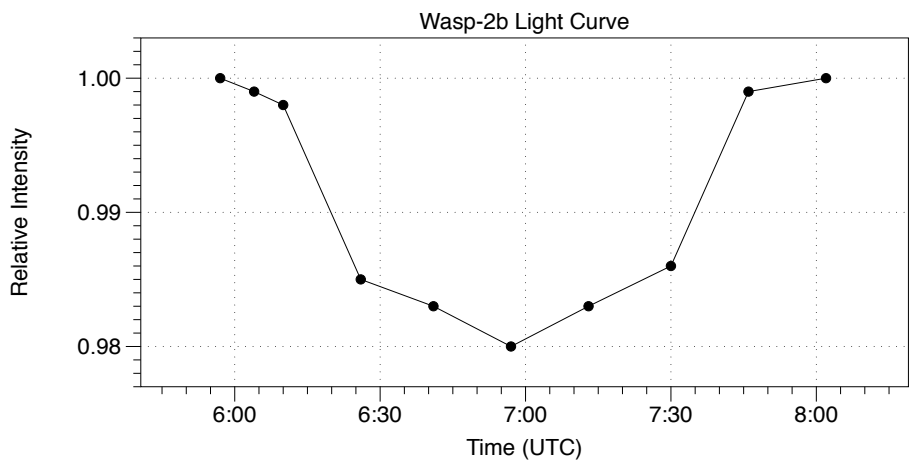

**Figure 4.** Sample light curve of WASP-2b

**Light Curve Analysis**

The dip in brightness observed during an exoplanet transit event is approximately related to the radii of the planet and star by

$$\Delta F \approx \left(\frac{R_p}{R_*}\right)^2 \quad (1)$$

where $F$ is the flux measured from the star, $\Delta F$ is the observed change in flux during transit, $R_p$ is the planet radius and $R_*$ is the stellar radius [5]. Represented visually as a light curve, we can geometrically analyse the results to help determine a number of characteristics about the exoplanet system. However, a good understanding of the host star is first required to accomplish this, and can be achieved by way of stellar classification.

*Stellar classification*

If you were to plot the luminosity (intrinsic brightness) vs. the colour (surface temperature) of many stars, the resulting scatter plot would show some interesting sequences. Known as the Hertzsprung-Russell diagram, this graph provides a visual representation of stellar populations and evolution (see Fig. 5). The most distinctive band of stars on the graph is known as the main-sequence. It can be seen that the luminosity of main-sequence stars increases with mass and is given by the mass–luminosity relation

$$\frac{L_*}{L_\odot} = \left(\frac{M_*}{M_\odot}\right)^4 \quad (2)$$

where $L_*$ and $M_*$ are the luminosity and mass of the star, and $L_\odot$ and $M_\odot$ are the luminosity and mass of the Sun. By conducting spectral observations, astronomers can determine the stellar luminosity, and by using known values of the Sun, the mass of the observed star can be determined. Previous observations of WASP-2 have yielded a luminosity of $1.9 \times 10^{26} \pm 0.2$ W [2]. Using this figure with equation (2), its mass was found to be $0.84 \pm 0.10\ M_\odot$.

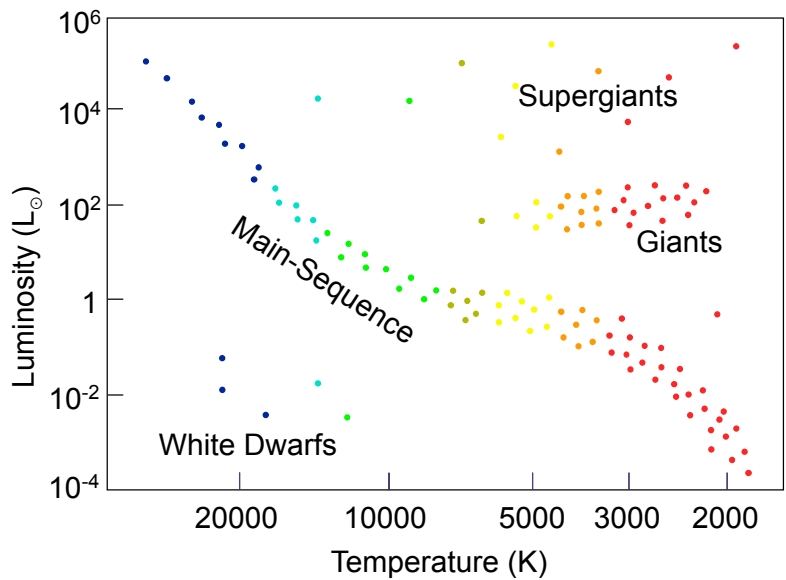

**Figure 5.** Scatter graph of the Hertzsprung-Russell diagram

*Geometric analysis*

One of the easiest parameters to determine upon the discovery of an exoplanet is its orbital period, $P$. The orbital period can be deduced from the timing of two mid-transit events, and is related to the semi-major axis of the orbit and the total mass of the two bodies via Kepler's third law,

$$\frac{a^3}{P^2} = \frac{G(M_* + M_p)}{4\pi^2} \tag{3}$$

where $M_*$ and $M_p$ are the mass of stellar and exoplanet mass respectively, $G$ is the gravitational constant, and $a$ is the semi-major axis. Thanks to multiple observations of WASP-2b, its orbital period is known to be $2.152226 \pm 0.000004$ days [2]. Assuming $M_* \gg M_p$ and a circular orbit, the semi-major axis of WASP-2b was found to be $0.031 \pm 0.001$ AU.

The orbital period and semi-major axis are also proportional to the transit duration, which in turn is dependent on the impact parameter, $b$. The impact parameter is the projected difference between the center of the stellar disc and the center of the planet disc at mid-transit, as seen by the observer (see Fig.6a). The impact parameter can be given by,

$$b = a \cos i \tag{4}$$

From Fig. 6b, it can be seen that the distance the planet travels from the center of the stellar disc to the edge of contact is length, $l$. Letting the impact parameter, $b$ equal to $a \cos i$, this length can be expressed in terms of the inclination, the semi-major axis and radii of the star and planet,

$$l = \sqrt{(R_* + R_p)^2 - a^2 \cos^2 i} \tag{5}$$

The length the planet travels across the star is $2l$, as seen by the observer. Fig. 6c shows that when the planet moves from point A to point B, it subtends an angle, $\alpha$. Assuming a circular orbit of $2\pi$ radians and the stellar radius is much larger than the exoplanet radius, the transit duration, $t$, can be given by,

$$t = \frac{P}{\pi} \sin^{-1}\left(\frac{\sqrt{(R_* + R_p)^2 - a^2 \cos^2 i}}{a}\right) \tag{6}$$

If we assume an edge-on orbit (90° angle of inclination) and $a \gg R_* \gg R_p$, equation 6 can be rewritten as

$$t \approx \frac{P R_*}{\pi a} \tag{7}$$

Inspection of the plotted light curve reveals that the transit took $1.8 \pm 0.2$ hours and caused a fractional dip in brightness of $0.020 \pm 0.005$. Using this data, the radius of the star was found (from equation 7) to be $0.73 \pm 0.10\ R_\odot$ and the radius of the planet (from equation 1) $1.05 \pm 0.27\ R_{\text{Jupiter}}$.

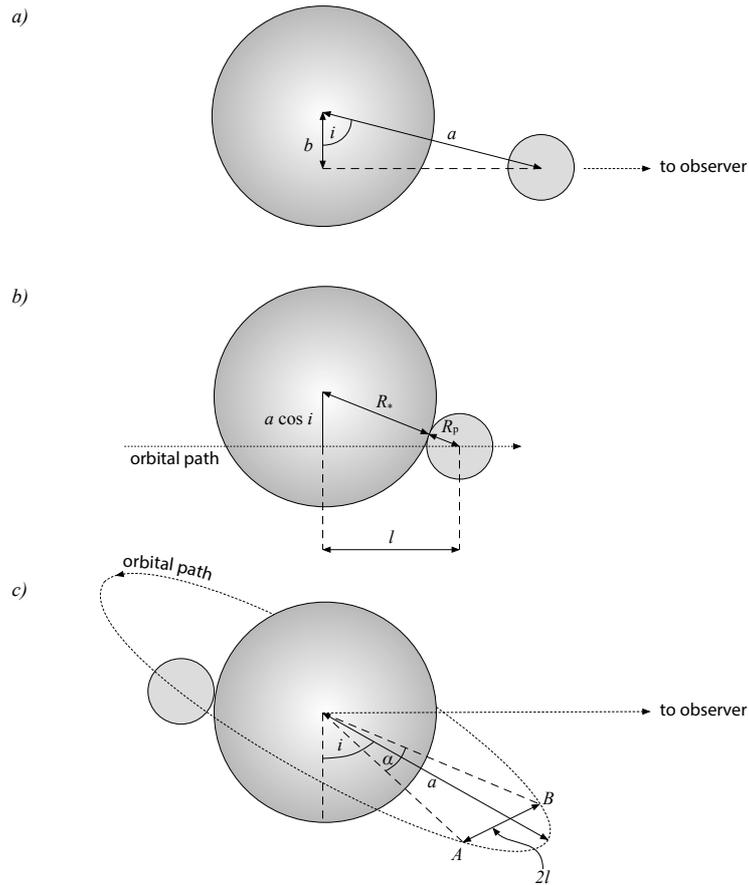

**Figure 6.** Geometry showing the impact parameter, *b* (part a) and the length, *l* (part b), which is related to the angle, $\alpha$ (part c), used to derive the transit duration (Adapted from [6, 7])

*Habitable zone*
The distance an exoplanet orbits its host star is an important factor in determining its ability to sustain liquid water, a precursor for life, as we know it. Given Earth lies within a region that's not too cold or too hot for liquid water to exist, our solar system is used as a template when classifying the habitable zone of another system (See Fig.7).

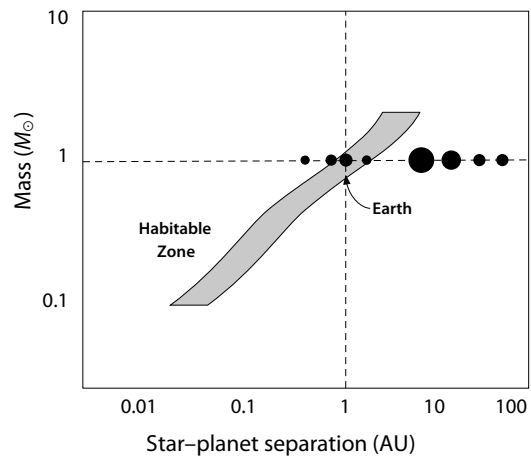

**Figure 7.** Position of the habitable zone for main-sequence stars, relative to their stellar mass

The size and location of the habitable zone is directly related to the luminosity of the stars. The extent of our own habitable zone is debated amongst scientist, but popular models place it at a range of 0.94 to 1.72 AU from the Sun [8]. Using the inverse square law, the range of a habitable zone for different stars can be given by,

$$d_{\text{inner}} = 0.94 \sqrt{\frac{L_*}{L_\odot}} \text{ AU}$$
$$d_{\text{outer}} = 1.72 \sqrt{\frac{L_*}{L_\odot}} \text{ AU} \tag{7}$$

where $L_*$ and $L_\odot$ is the luminosity of the star and Sun respectively, and $d$ is the distance to the habitable zone edge, in astronomical units. From our earlier results, WASP-2b was found to orbit its host at a distance of 0.031 ± 0.001 AU, but the system's habitable zone ranges approximately from 0.66 to 1.2 AU, suggesting WASP-2b is too close to its host star to be considered habitable.

A summary of the characteristics calculated above and accepted values [2] is displayed in the table below.

| Parameter | Calculated | Accepted Values |
| --- | --- | --- |
| WASP-2 Mass | 0.84 ± 0.10 Solar Masses | 0.85 ± 0.10 Solar Masses |
| WASP-2 Radius | 0.73 ± 0.10 Solar Radii | 0.78 ± 0.06 Solar Radii |
| WASP-2b Radius | 1.05 ± 0.27 Jupiter Radii | 0.96 ± 0.3 Jupiter Radii |
| Orbital Distance | 0.031 ± 0.001 AU | 0.031 ± 0.001 AU |

**Discussion**
The activity described in this paper represents a basic approach to plotting a light curve from a transiting exoplanet. As such, the calculated results represent rough approximations, caused by a number of limiting factors, such as imperfections in the images, low number of reference stars in the field of view, and oversimplification of the light curve geometry. Despite this, the results compared well with currently accepted values. To help students learn the value of scientific method, they should produce their own results and compare them to accepted values of WASP-2b, shown above or from the SuperWASP site (www.superwasp.org/).

This activity has been performed in a number of workshops for advanced secondary school students and university undergraduates. The workshops involved a 1-hour introductory session on exoplanets including a practical simulation, similar to the demonstrations performed in [9, 10]. Students were then allocated 2 hours to individually produce and analyse the transit light curves, and later performed a presentation on their findings to the class. Student reactions were positive, as they seemed not only to enjoy the activity, but also learned something new.